\documentclass[fleqn,10pt,twocolumn,a4paper]{SICE_ISCS}

\title{Communication Performance Analysis of \\
Sampled-Data $H^{\infty}$ Optimal Coupling Wave Canceler}

\author{Hampei Sasahara${}^{1\dagger}$, Masaaki Nagahara${}^{1}$, Kazunori Hayashi${}^{1}$, and Yutaka Yamamoto${}^{1}$}
\speaker{Hampei Sasahara}

\affils{${}^{1}$ Graduate School of Informatics, Kyoto University, Kyoto, Japan\\
(E-mail:
sasahara.h@acs.i.kyoto-u.ac.jp,
nagahara@i.kyoto-u.ac.jp,
kazunori@i.kyoto-u.ac.jp,
yy@i.kyoto-u.ac.jp
)
}
\abstract{%
In this manuscript, we propose a design method of digital filters which cancel coupling waves
generated in single-frequency full-duplex wireless relay stations by using the sampled-data $H^{\infty}$ control theory.
Simulation results show effectiveness of the proposed method to communication performance from a base station to a terminal.
 }

\keywords{%
Wireless communication, Coupling waves, Sampled-data $H^{\infty}$ optimization, Relay station
}

\begin{document}

\maketitle


\section{Introduction}
In wireless communications,
relay stations are used to relay radio signals between radio stations that cannot
directly communicate with each other due to the signal attenuation.
On the other hand, {\it single-frequency network} is expected in which signals with the same carrier frequency are transmitted through communication network,
in order to efficiently utilize the scarce bandwidth due to the limitation of frequency resources.
Then, a problem of {\it coupling waves} arises in a full-duplex relay station in a single-frequency network \cite{Jain}.

For this problem,
the authors have proposed a design method of digital coupling wave cancelers
that take account of suppression of coupling wave effects including the intersample behavior
and stability of the closed-loop system simultaneously \cite{SSHR_JCMSI14}.
The communication performance of the obtained filter is estimated as a bit error rate (BER) through a simulation result in this article,
while the performance of the designed canceler has been evaluated as time-domain reconstruction performance in~\cite{SSHR_JCMSI14}.

\section{System model}
\begin{figure}[t]
\includegraphics[width = \linewidth]{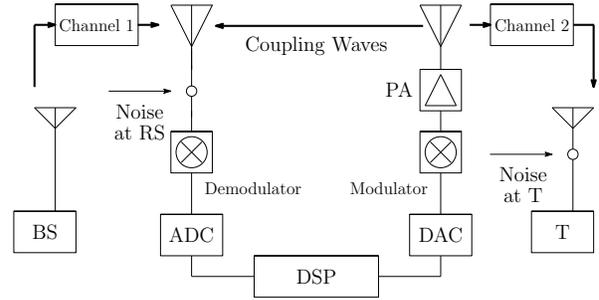}
\caption{The system model from BS to T}
\label{BStoTmodel}
\end{figure}

In the study, it is assumed that we use the amplify-and-forward (AF) protocol,
which is implemented with ease, for the relay station (RS) \cite{Nabar04}.
Fig.~\ref{BStoTmodel} depicts the model of the system from a base station (BS) to a terminal (T),
where ADC is an analog-to-digital converter and DAC is a digital-to-analog converter respectively.
The power amplifier is denoted by PA and DSP represents a digital signal processor.
The AF type relay station is located between the base station and the terminal.
Note that coupling waves occur between the antennas at RS and noises are added at the receiving antennas of RS and T.
Finally, we consider the quadrature amplitude modulation for the modulation protocol.

\begin{figure}[t]
\includegraphics[width = \linewidth]{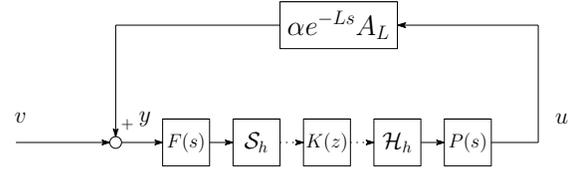}
\caption{The block diagram of the relay station}
\label{relaymodel}
\end{figure}

After some transforms from Fig.~\ref{BStoTmodel},
we obtain the model of the relay station shown in \cite{SSHR_JCMSI14}.
In Fig.~\ref{relaymodel}, $v$ is the received baseband signal,
$u$ is the output signal after cancelation of coupling waves.
Note that here we do not consider the noise at RS.
$F(s)$ is an anti-aliasing analog filter with the ideal sampler ${\mathcal S}_h$ with sampling period $h$
and $P(s)$ is a post analog low-pass filter with the zero-order hold ${\mathcal H}_h$ with sampling period $h$.
The relay gain and the characteristic of coupling wave path are denoted by $\alpha e^{-Ls}A_L$.
Here, $\alpha$ is the product of the relay gain and the attenuation of coupling wave,
$L$ is the delay time of the coupling wave path,
and $A_L$ is the two-dimensional rotation matrix with its angle $-2\pi fL$,
where $f$ is the carrier frequency.
$K(z)$ denotes a digital controller, which we design for coupling wave canceler.

\begin{figure}[t]
\includegraphics[width = \linewidth]{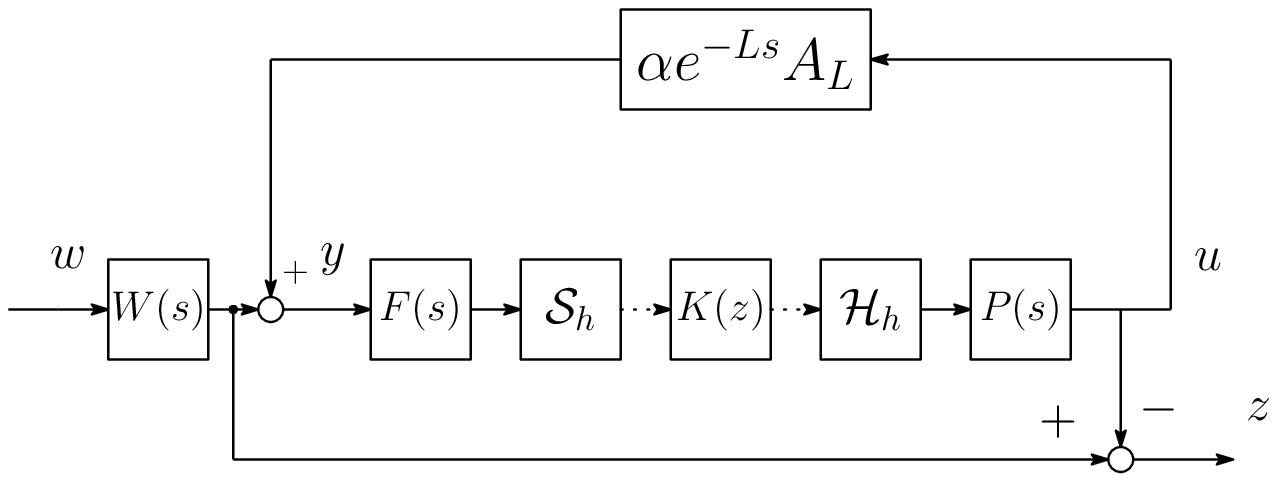}
\caption{The block diagram for the digital controller design}
\label{BlkDia}
\end{figure}

For the obtained system described in Fig.~\ref{relaymodel}, we find the digital controller, $K(z)$, that stabilizes the feedback system
and also minimizes the effect of coupling wave, $z=v-u,$ for any $v$.
To obtain a reasonable solution,
we restrict the input continuous-time signal $v$ to the following set
\begin{eqnarray}
WL^2 := \{v=Ww:w\in L^2, \|w\|_2 =1\},
\end{eqnarray}
where $W(s)$ is a continuous-time, linear, and time-invariant system with a real-rational stable, and strictly proper transfer function, and $L^2$ is the Lebesgue space consisting of all square integrable real functions on $[0, \infty)$ endowed with $L^2$ norm $\| \cdot \|_2$.
Fig.~\ref{BlkDia} shows the block diagram for the digital controller design.

Let $T_{zw}$ be the system from $w$ to $z$.
Here we formulate our design problem to find a digital controller $K(z)$ that minimizes $\|T_{zw}\|_{\infty}$.
This is a standard sampled-data $H^{\infty}$ control problem, and can be efficiently solved by the fast sample fast hold (FSFH) approximation \cite{Chen}.
Note that if there exists a controller $K(z)$ that minimizes $\|T_{zw}\|_{\infty}$,
then the feedback system is stable and the effect of coupling wave interference $z=v-u$ is bounded by the minimum $H^{\infty}$ norm.

\section{Simulation}
We here analyze the communication performance of the proposed canceler with computer simulation.
Simulation parameters are shown in Table~1.
Note that the sampling period and the carrier frequency are normalized.
The attenuation rate of the coupling wave channel is $0.15$,
the delay time $L=1$,
and the anti-alias analog filter $F(s)=I$.
The post filter $P(s)$ is modeled by
\begin{eqnarray}
P(s) = \frac{1}{0.001s+1}I.
\end{eqnarray}
The frequency characteristic of the input signals is modeled by the following real-rational transfer function
\begin{eqnarray}
W(s) = \frac{1}{2s+1}I.
\end{eqnarray}
The proposed method does not employ the information of the channels from BS to RS and from RS to T for the filter design.
To concentrate effect of suppressed coupling waves, we assume that the characteristic of the channel from BS to RS, the channel 1 in Fig.~\ref{BStoTmodel},
is modeled as $1$ (identity),
and the characteristic of the channel from RS to T, the channel 2,
is modeled as a scalar gain $\beta$.

\begin{table}[t]
\caption{Simulation parameters}
\begin{center}
\begin{tabular}{|c|c|}\hline
modulation & binary phase shift keying \\\hline
sampling period & $1$ [sec] \\\hline
carrier frequency& $10000$ [Hz]  \\\hline
transmission pulse & rectangular (time-domain) \\\hline
symbol period& $2$ [sec]\\\hline
relay gain & $60$ [dB] \\\hline
noise & white Gaussian \\\hline
signal power at BS& $0$ [dBm]\\\hline
noise power at RS& $-5$ [dBm]\\\hline
noise power at T& $-2$ [dBm]\\\hline
FSFH number & $16$ \\\hline
symbol number & $10000$ \\\hline
\end{tabular}
\end{center}
\end{table}

\begin{figure}[t]
\centering
\includegraphics[width = 0.9\linewidth]{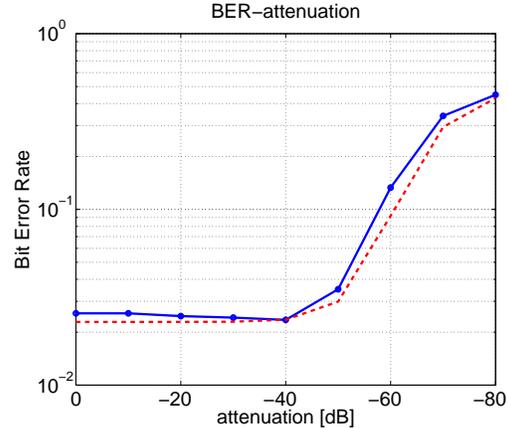}
\caption{BER at T vs $\beta,$ which is the attenuation of the RS-T channel: with designed filter cancelation (solid line) and perfect cancelation  (dashed line) }
\label{01}
\end{figure}

Fig.~\ref{01} shows the BER at T vs $\beta$.
The solid line represents the BER vs $\beta$ with the designed filter,
and the dashed line represents the ideal curve with perfect cancelation of coupling wave effect.
This shows the designed filter almost removes the effect of coupling waves in the sense of communication performance.
In this figure, the smaller $\beta$ is, the worse the BER goes.
The reason is that the signal to noise ratio becomes low between the transmission signal from the relay station and the noise at the terminal
when $\beta$ is small.

\section{Conclusions}
In this manuscript, we have proposed filter design for coupling wave cancellation based on the sampled-data $H^{\infty}$ control theory.
The simulation result has been shown to illustrate the effectiveness of the proposed canceler.

\section*{Acknowledgement}
This research is supported in part by the JSPS Grant-in-Aid for Scientific Research (B) No.~24360163
and (C) No.~24560543, Grant-in-Aid for Scientific Research on Innovative Areas No.~26120521,
and an Okawa Foundation Research Grant.



\begin{thebibliography}{9}
\bibitem{Jain}
M. Jain {\it et al.,}
``Practical, real-time, full duplex wireless,''
{\it Proc. of 17th Annual International Conference on Mobile Computing and Networking (MobiCom),} 
pp.~301--312, 2011.

\bibitem{SSHR_JCMSI14}
H.~Sasahara, M.~Nagahara, K.~Hayashi, and Y.~Yamamoto,
``Digital cancelation of self-interference for single-frequency full-duplex relay stations via sampled-data $H^{\infty}$ control,''
{\it SICE JCMSI,} 2015. {\it (submitted)}

\bibitem{Nabar04}
R.~U.~Nabar, H.~Bolcskei, and F.~W.~Kneubuhler,
``Fading relay channels: performance limits and space-time signal design,''
{\it Selected Areas in Communications, IEEE Journal on,} Vol.~22, no.~6, pp.~1099--1109, 2004.

\bibitem{Chen}
T.~Chen and B.~A.~Francis,
{\it Optimal Sampled-Data Control Systems,} Springer, 1995.

\end{thebibliography}
\end{document}